# The Moral Consideration of Artificial Entities: A Literature Review

Jamie Harris: Sentience Institute, jamie@sentienceinstitute.org
Jacy Reese Anthis: Sentience Institute & Department of Sociology, University of Chicago

# Abstract

Ethicists, policy-makers, and the general public have questioned whether artificial entities such as robots warrant rights or other forms of moral consideration. There is little synthesis of the research on this topic so far. We identify 294 relevant research or discussion items in our literature review of this topic. There is widespread agreement among scholars that some artificial entities could warrant moral consideration in the future, if not also the present. The reasoning varies, such as concern for the effects on artificial entities and concern for the effects on human society. Beyond the conventional consequentialist, deontological, and virtue ethicist ethical frameworks, some scholars encourage "information ethics" and "social-relational" approaches, though there are opportunities for more in-depth ethical research on the nuances of moral consideration of artificial entities. There is limited relevant empirical data collection, primarily in a few psychological studies on current moral and social attitudes of humans towards robots and other artificial entities. This suggests an important gap for social science research on how artificial entities will be integrated into society and the factors that will determine how the interests of sentient artificial entities are considered.

# Introduction

Humans could continue to exist for many centuries on earth or expand to far greater interstellar populations in the coming millennia (e.g. very approximately $10^{38}$ human lives per century if we colonize the Virgo Supercluster (Bostrom 2003)).  Artificial entities could be less complex and require less energy than biological humans, so there could be even more artificial entities in the future. Even if one assigns a small likelihood that artificial entities will propagate, their potentially astronomical numbers raise an important question of moral consideration.

Policy-makers have begun to engage with this question. A 2006 paper commissioned by the U.K. Office of Science argued that robots could be granted rights within 50 years (BBC 2006). South Korea proposed a "robot ethics charter" in 2007 (Kim 2010). Paro, a type of care robot in the shape of a seal, was granted a "koseki" (household registry) in Nanto, Japan in 2010 (Robertson 2014). The European Parliament passed a resolution in 2017 suggesting the creation of "a specific legal status for robots in the long run, so that at least the most sophisticated autonomous robots could be established as having the status of electronic persons" (European Parliament Committee on Legal Affairs 2017). In the same year, a robot named Sophia was granted citizenship in Saudi Arabia (Hanson Robotics 2018) and a chatbot on the messaging app Line, named Shibuya Mirai, was granted residence by the city of Tokyo in Japan (Microsoft Asia News Center 2017).

Policy decisions relating to the rights of artificial entities have been reported in the media (Browne 2017; Maza 2017; Reynolds 2018; Weller 2020), discussed by the public,[1] and critiqued by academics (Open Letter 2018). The moral consideration of artificial entities has also been explored extensively in science fiction (McNally and Inayatullah 1988, p. 128; Petersen 2007, pp. 43-4; Robertson 2014, p. 573-4; Inyashkin 2016; Kaminska 2016; Arnold and Gough 2017; Gunkel 2018a, pp. 13-8; Hallqvist 2018; Kunnari 2020). There have been some relevant advocacy efforts, arguably including by the robot Sophia (Create Digital 2018), but also two groups of humans: People for the Ethical Treatment of Reinforcement Learners (PETRL 2015) and The American Society for the Prevention of Cruelty to Robots (Anderson 2015).

Scholars often conclude that artificial entities with the capacity for positive and negative experiences (i.e. sentience) will be created, or are at least theoretically possible (see, for example, Thompson 1965; Aleksander 1996; Buttazzo 2001; Blackmore 1999; Franklin 2003; Harnad 2003; Holland 2007; Chrisley 2008; Seth 2009; Haikonen 2012; Bringsjord et al. 2015; Angel 2019) and an informal survey of Fellows of the American Association for Artificial Intelligence suggested that many were open to this possibility (McDermott 2007). Tomasik (2011), Bostrom (2014), Gloor (2016a), and Sotala and Gloor (2017) argue that the insufficient moral consideration of sentient artificial entities, such as the subroutines or simulations run by a future superintelligent AI, could lead to astronomical amounts of suffering. Kelley and Atreides (2020) have already proposed a "laboratory process for the assessment and ethical treatment of Artificial General Intelligence systems that could be conscious and have subjective emotional experiences."

There has been limited synthesis of this literature to date. Gunkel (2018a) provides the most thorough review to set up his argument about "robot rights," categorizing contributions into four modalities: "Robots Cannot Have Rights; Robots Should Not Have Rights," "Robots Can Have Rights; Robots Should Have Rights," "Although Robots Can Have Rights, Robots Should Not Have Rights," and "Even if Robots Cannot Have Rights, Robots Should Have Rights." Gunkel critiques each of these perspectives, advocating instead for "thinking otherwise" via deconstruction of the questions of whether robots can and should have rights. Bennett and Daly (2020) more briefly summarize the literature on these two questions, adding a third: "*will* robots be granted rights?" They focus on legal rights, especially legal personhood and intellectual property rights. Tavani (2018) briefly reviews the usage of "robot" and "rights," the criteria necessary for an entity to warrant moral consideration, and whether moral agency is a prerequisite for moral patiency, in order to explain a new argument that social robots warrant moral consideration.

However, those reviews have not used systematic methods to comprehensively identify relevant publications or quantitative methods of analysis, making it difficult to extract general trends and themes.[2] Do scholars tend to believe that artificial entities warrant moral consideration? Are views split along geographical and disciplinary lines? Which nations, disciplines, and journals most frequently provide contributions to the discussion? Using a systematic search methodology, we address these questions, provide an overview of the literature, and suggest opportunities for further research. Common in social science and clinical research (see, for example, Higgins and Green 2008; Campbell Collaboration 2014), systematic reviews have recently been used in philosophy and ethics research (Nill and Schibrowsky 2007; Mittelstadt 2017; Saltz and Dewar 2019).

Previous reviews have also tended to focus on "robot rights." Our review has a broader scope. We use the term "artificial entities" to refer to all manner of machines, computers, artificial intelligences, simulations, software, and robots created by humans or other entities. We use the phrase "moral consideration" of artificial entities to collectively refer to a number of partly overlapping discussions: whether artificial entities are "moral patients," deserve to be included in humanity's moral circle, should be granted "rights," or should otherwise be granted consideration. Moral consideration does not necessarily imply the attribution of intrinsic moral value. While not the most common,[3] these terms were chosen for their breadth.

---

[1] See, for example, the comments on Barsanti (2017).

[2] Vakkuri and Abrahamsson (2018) use a systematic methodology to examine key words. However, only 83 academic papers are examined, with "rights" only being returned as a key word in two of the articles and "moral patiency" in another three. As such, it is not comprehensive.

# Methodology

Four scientific databases (Scopus, Web of Science, ScienceDirect, and the ACM Digital Library) were searched systematically for relevant items in August and September 2020. Google Scholar was also searched, since this search engine is sometimes more comprehensive, particularly in finding the grey literature that is essential to cataloguing an emerging field (Martín-Martín et al. 2019).

Given that there is no single, established research field examining the moral consideration of artificial entities, multiple searches were conducted to identify relevant items; a total of 2,692 non-unique items were screened for inclusion (see Table 1). After exclusions (see criteria below) and removal of duplicates, 294 relevant research or discussion items were included (see Table 2).

Table 1: Initial results returned for screening, by search terms and search location

| Search category | Search term | Scopus | Web of Science | Science Direct | ACM Digital Library | Google Scholar |
|---|---|---|---|---|---|---|
| "rights" | "robot rights" | 42 | 22 | 13 | 29 | 939 |
| | "rights for robots" | 4 | 3 | 33 | 7 | 179 |
| | "machine rights" OR "rights for machines"[a] | 35 | 6 | 36 | 3 | 267 |
| | "artificial intelligence rights" OR "rights for artificial intelligence" | 2 | 1 | 13 | 0 | 59 |
| "moral" | ("moral consideration" OR "moral concern") AND (robots OR machines OR "artificial intelligence") | 54 | 12 | 529 | 102 | 8,690 |
| | ("moral circle" OR "moral expansiveness") AND (robots OR machines OR "artificial intelligence") | 4 | 2 | 13 | 5 | 420 |
| | ("Moral patient" OR "moral patients" OR "moral patiency") AND (robots OR machines OR "artificial intelligence") | 25 | 11 | 28 | 52 | 1,290 |
| "suffering" | "suffering subroutines" | 0 | 0 | 0 | 0 | 18 |
| | ("mind crime" OR "mindcrime") AND simulations | 0 | 0 | 0 | 2 | 82 |
| | ("astronomical suffering" OR "suffering risks") | 11 | 5 | 46 | 4 | 277 |

[3] See "Focus and terminology." For examples of their use, see Küster and Swiderska (2020) and Coeckelbergh (2010b).

[a] The Google Scholar search for "Machine rights" OR "rights for machines" included the additional operator -Kathrani, because the search otherwise returned a large number of webpages that all referred back to a single talk.

Table 2: Included items, by search terms and search location

| Search category | Search term | Scopus | Web of Science | Science Direct | ACM Digital Library | Google Scholar |
|---|---|---|---|---|---|---|
| "rights" | "robot rights" | 20 | 13 | 4 | 18 | 70 |
| | "rights for robots" | 2 | 1 | 6 | 4 | 50 |
| | "machine rights" OR "rights for machines" | 6 | 5 | 2 | 1 | 23 |
| | "artificial intelligence rights" OR "rights for artificial intelligence" | 2 | 1 | 2 | 0 | 9 |
| "moral" | ("moral consideration" OR "moral concern") AND (robots OR machines OR "artificial intelligence") | 20 | 9 | 6 | 40 | 82 |
| | ("moral circle" OR "moral expansiveness") AND (robots OR machines OR "artificial intelligence") | 2 | 0 | 5 | 2 | 27 |
| | ("Moral patient" OR "moral patients" OR "moral patiency") AND (robots OR machines OR "artificial intelligence") | 14 | 7 | 4 | 27 | 75 |
| "suffering" | "suffering subroutines" | 0 | 0 | 0 | 0 | 9 |
| | ("mind crime" OR "mindcrime") AND simulations | 0 | 0 | 0 | 1 | 14 |
| | ("astronomical suffering" OR "suffering risks") | 2 | 2 | 1 | 0 | 8 |

For the database searches, the titles and abstracts of returned items were reviewed to determine relevance. For the Google Scholar searches, given the low relevance of some returned results, review was limited to the first 200 results, similar to the approach of Mittelstadt (2017).

Common reasons for exclusion were that the item:

- Did not discuss the moral consideration of artificial entities (e.g. discussed whether artificial entities could be moral agents but not whether they could be moral patients[4]),
- Mentioned the topic only very briefly (e.g. only as thought-provoking issue adjacent to the main focus of the article), or

[4] See, for example, Wallach at al. (2008). We recognize that some moral frameworks may see moral agency as an important criterion affecting moral consideration (see, for example, Wareham 2013; Laukyte 2017). However, this criterion seems less directly relevant and including it in this literature review would have substantially widened the scope. Evaluations of agency and patiency may be correlated, but artificial entities may be assigned high agency alongside low patiency (Gray et al. 2007; Akechi et al. 2018). Lee et al. (2019) found that manipulations of patiency significantly affected perceived agency but that the reverse was not true.

- Were not in the format of an academic article, book, conference paper, or peer-reviewed magazine contribution (e.g. they were published as a newspaper op-ed or blog post).

The findings are analyzed qualitatively and discussed in the sections below. Results are also categorized and scored along the following dimensions:

- Categories of the search terms that identified each item, which reflect the language used by the authors; the three categories used are “rights,”, “moral,” and “suffering” searches,
- Categories of academic disciplines of the lead author of each included item,
- Categories of primary frameworks or moral schemas used, and
- A score representing the author’s position on granting moral consideration to artificial entities on a scale from 1 (argues forcefully against consideration, e.g. suggesting that artificial beings should never be considered morally) to 5 (argues forcefully for consideration, e.g. suggesting that artificial beings deserve moral consideration now).

In addition to the discussion below, Appendix A includes a summary of each item and the full results of the categorization and scoring analyses.

# Results

## Descriptive statistics

Included items were published in 105 different journals. Four journals published more than five of the included items; *Ethics and Information Technology* (9% of items), *AI and Society* (4%), *Philosophy and Technology* (2%), and *Science and Engineering Ethics* (2%). Additionally, 15% of items were books or chapters (only one book focused on this topic was identified, Gunkel 2018a),[5] 13% were entries in a report of a conference, workshop, or symposium (often hosted by the Association for Computing Machinery or Institute of Electrical and Electronics Engineers), and 12% were not published in any journal, magazine, or book.

The included items were produced by researchers affiliated with institutions based in 43 countries. Only five countries produced more than 10 of the identified items: the United States (36% of identified items), the United Kingdom (15%), the Netherlands (7%), Australia (5%), and Germany (4%). According to Google Scholar, included items have been cited 5,992 times (excluding one outlier with 2,513 citations, Bostrom 2014); 41% of these citations are of items produced in the US.[6]

The oldest included item identified by the searches was McNally and Inayatullah (1988), though included items cited articles from as early as 1964 as offering relevant comments (Putman 1964; Stone 1974; Lehman-Wilzig 1981; Freitas 1985). The study of robot ethics (now called “roboethics” by some (Veruggio and Abney 2012, pp. 347-8)) grew in the early 2000s (Malle 2016). Levy (2009), Torrance (2013, p. 403), and Gunkel (2018c, p. 87) describe the moral consideration of artificial entities as a small and neglected sub-field. However, the results of this literature review suggest that academic interest in the moral consideration of artificial entities is growing exponentially (see Figure 1).

Figure 1: Cumulative total of included items, by date of publication.

[5] Gellers 2020 has subsequently been published.
[6] If Bostrom (2014) is included, then only 29% of citations were of items produced in the US, compared to 51% in the UK.

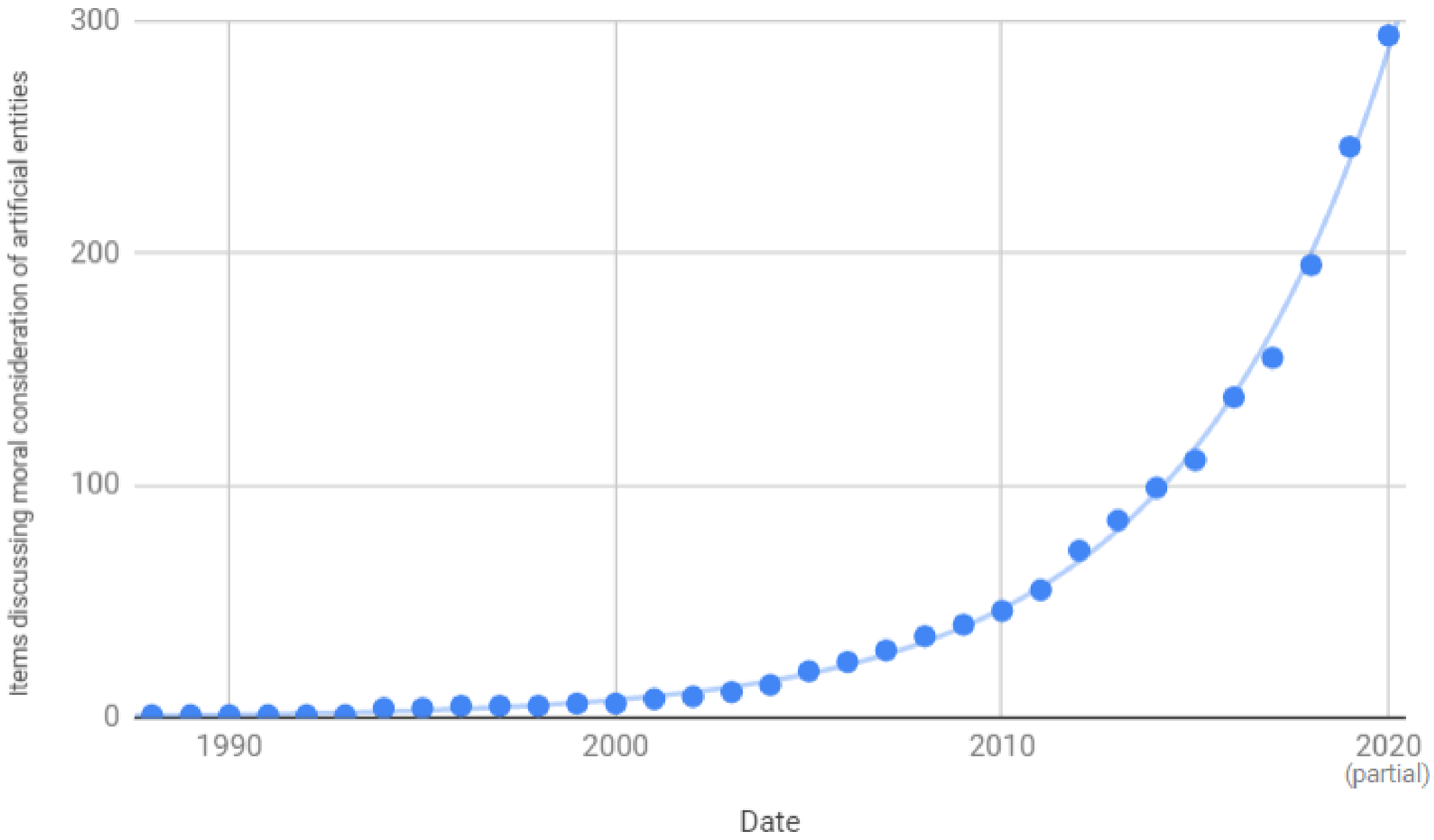

As shown in Table 3, the most common academic disciplines of contributing scholars are philosophy or ethics, law, computer engineering or computer science, and communication or media. We focus on the primary disciplines of scholars, rather than of publications, because so many of the publications are interdisciplinary.

Table 3: Items and citations by the academic discipline of the lead author.

| | Count and citations | | Percentages of total | |
|---|---|---|---|---|
| **Academic discipline** | **Count of items** | **Citation count (outlier excluded)[a]** | **Count of items** | **Citation count (outlier excluded)** |
| Philosophy or ethics | 82 | 2101 | 28% | 35% |
| Law | 49 | 1002 | 17% | 17% |
| Computer engineering or computer science | 30 | 473 | 10% | 8% |
| Communication or media | 28 | 740 | 10% | 12% |
| Psychology | 20 | 495 | 7% | 8% |
| Other social sciences | 18 | 111 | 6% | 2% |
| Other engineering | 12 | 200 | 4% | 3% |
| Cognitive science | 10 | 300 | 3% | 5% |
| Other humanities | 8 | 14 | 3% | 0% |
| Medical or biology | 6 | 111 | 2% | 2% |
| Information technology | 5 | 86 | 2% | 1% |
| Design | 5 | 27 | 2% | 0% |
| Robotics | 4 | 23 | 1% | 0% |
| History | 3 | 83 | 1% | 1% |
| Business | 3 | 17 | 1% | 0% |
| Political science | 3 | 13 | 1% | 0% |
| English literature or language | 3 | 0 | 1% | 0% |
| Future studies | 2 | 3 | 1% | 0% |
| Other or unidentifiable | 32 | 471 | 11% | 8% |
| **Total** | **294** | **5992** | **100%** | **100%** |

[a] The outlier is Bostrom (2014), with 2,513 citations, which is categorized under "philosophy or ethics." If this item is included, then the citation count for philosophy or ethics rises to 4614, which is 54% of the total.

As shown in Table 4, many scholars contributing to the discussion do not adopt a single, clear moral schema, focusing instead on legal precedent, empirical evidence of attitudes towards artificial entities, or simply summarizing the views of previous scholars (e.g. Weng et al. 2009, p. 267; Gray and Wegner 2012, pp. 125-30).

Table 4: Items and citations by the primary framework or moral schema used.

| | **Count and citations** | | **Percentages of total** | |
|---|---|---|---|---|
| **Primary framework or moral schema** | **Count of items** | **Citation count (outlier excluded)[a]** | **Count of items** | **Citation count (outlier excluded)** |
| Legal precedent | 34 | 906 | 12% | 15% |
| Social-relational | 23 | 606 | 8% | 10% |
| Consequentialist | 16 | 99 | 5% | 2% |
| Deontological | 15 | 213 | 5% | 4% |
| Information ethics | 5 | 1019 | 2% | 17% |
| Virtue ethicist | 5 | 28 | 2% | 0% |
| Not Applicable | 103 | 1641 | 35% | 27% |
| Mixture | 52 | 892 | 18% | 15% |
| Unclear | 26 | 277 | 9% | 5% |
| Other | 15 | 311 | 5% | 5% |
| **Total** | **294** | **5992** | **100%** | **100%** |

[a] The outlier is Bostrom (2014), with 2,513 citations, which was categorized as using a consequentialist moral framework. If this item is included, then the citation count for the consequentialist category rises to 2612, which is 31% of the total.

Many scholars use consequentialist, deontological, or virtue ethicist moral frameworks, or a mixture of these. These scholars defend various criteria as crucial for determining whether artificial entities warrant moral consideration. Sentience or consciousness seem to be most frequently invoked (Himma 2003; Torrance 2008; Basl 2014; Mackenzie 2014; Bostrom 2014; Tomasik 2014; Yampolskiy 2017; Johnson and Verdicchio 2018; Andreotta 2020; Mosakas 2020), but other proposed criteria include the capacities for autonomy (Calverley 2011; Neely 2014; Gualeni 2020), self-control (Wareham 2013), rationality (Laukyte 2017), integrity (Gualeni 2020), dignity (Bess 2018), moral reasoning (Malle 2016), and virtue (Gamez et al. 2020).

Some of the most influential scholars propose more novel ethical frameworks. Coeckelbergh (2010a, 2010b, 2014, 2018, 2020) and Gunkel (2013, 2014, 2015, 2018a, 2018b, 2018c, 2018d, 2019a, 2019b, 2020), encourage a social-relational framework to discuss the moral consideration of artificial entities. This approach grants moral consideration on the basis of how an entity "is treated in actual social situations and circumstances" (Gunkel 2018, p. 10). Floridi (1999, 2002, 2005) encourages "information ethics," where "[a]ll entities, qua informational objects, have an intrinsic moral value." Though less widely cited, Danaher's (2020) theory of "ethical behaviorism" and Tavani's (2018) discussion of "being-in-the-technological-world" arguably also offer alternative moral frameworks for assessing whether artificial entities warrant moral consideration.

# Focus and terminology

Definitions of the widely-used term "robot" are varied and often vague (Lin et al. 2011, pp. 943-4; Robertson 2014, p. 574; Tavani 2018, pp. 2-3; Gunkel 2018a, pp. 14-26; Beno 2019, pp. 2-3). It can be defined broadly, such as "a machine that resembles a living creature in being capable of moving independently (as by walking or rolling on wheels) and performing complex actions (such as grasping and moving objects)" (Merriam-Webster 2008). More narrowly, to many people, the term robot implies a humanoid appearance, or at least humanoid functions and behaviors (Brey and Søraker 2009; Leenes and Lucivero 2014; Rademeyer 2017). This terminology seems suboptimal, given that the forms of artificial sentience that seem most at risk of experiencing intense suffering on a large scale in the long-term future may not have humanoid characteristics or behaviors; they may even exist entirely within computers, not having any embodied form, human or otherwise.[7] Other terms used by scholars include "artificial beings" (Gualeni 2020), "artificial consciousness" (Basl 2013b), "artificial entities" (Gunkel 2015), "artificial intelligence" (Ashrafian 2015b), "artificial life" (Sullins 2005), "artificial minds" (Jackson Jr 2018), "artificial person" (Michalski 2018), "artificial sentience" (Ziesche and Yampolskiy 2019), "machines" (Church 2019), "automata" (Miller 2015), computers (Drozdek 1994), "simulations" (Bostrom 2014), and "subroutines" (Winsby 2013). Alternative adjectives such as "synthetic," "electronic," and "digital" are also sometimes used to replace "artificial."[8]

Relevant discussion has often focused on the potential "rights" of artificial entities (Tavani 2018, pp. 2-7; Gunkel 2018a, pp. 26-33). There has been some debate over whether "rights" is the most appropriate term, given its ambiguity and that legal and moral rights are each only one mechanism for moral consideration (Kim and Petrina 2006, p. 87; Tavani 2018, pp. 4-5; Cappuccio et al. 2020, p. 4). Other scholars consider whether artificial entities can be "moral patients," granted "moral consideration," or included in the "moral circle" (Küster and Świderska 2016; Cappuccio et al. 2020; Danaher 2020). Some scholars use terminology that focuses on the suffering of specific forms of artificial sentience: "mind crime" against simulations (Bostrom 2014), "suffering subroutines" (Tomasik 2011), or "risks of astronomical future suffering" (Tomasik 2011) and the derivative term "s-risks."

[7] Presumably, sentient subroutines (as discussed in Tomasik 2011) would not have humanoid shape, though some sentient simulations could have a humanoid shape in their simulated environment.

[8] In the identified results, these adjectives tended to be used alongside "artificial" (see, for example, San José et al. 2016), though this may reflect the search terms used in this literature review. These adjectives were not included in the search terms because initial exploration suggested that the vast majority of returned results were irrelevant to the focus of this literature review.

There were more items found by “rights” or “moral” than “suffering” search terms (see Table 5). Although 31% of the items identified by “rights” search terms were also identified by “moral” search terms, the results from the “suffering” search terms were more isolated, with only 12% of those items also being identified by “rights” or “moral” search terms. Additionally, excluding one outlier—Bostrom (2014)[9]—items identified via the “suffering” search terms had a lower average citation count (8) than items identified via “moral” (24) or “rights” (20) search terms. If the outlier is included, then the average for the suffering search terms is over ten times larger (108), and these items comprise 32% of the total citations (see Appendix A).

Table 5: Items and citations by search term category.

| | Count and citations | | Percentages of total | |
|---|---|---|---|---|
| **Search term category** | **Count of items** | **Citation count (outlier excluded)** | **Count of items** | **Citation count (outlier excluded)** |
| “Rights” search terms | 146 | 2938 | 50% | 49% |
| “Moral” search terms | 171 | 4071 | 58% | 68% |
| “Suffering” search terms | 25 | 187 | 9% | 3% |
| **Total** | **294** | **5992** | **100%** | **100%** |

The terminology used varied by the authors’ academic discipline and moral framework (see Appendix A). For example, the items by legal scholars were mostly identified by “rights” search terms (80%) while the items by psychologists were mostly identified by “moral” search terms (90%). In the “other or unidentifiable” category, 44% were identified via “suffering” search terms; these contributions were often by the Center on Long-Term Risk and other researchers associated with the effective altruism community.[10] An unusually high proportion of “consequentialist” items were identified by “suffering” search terms (50%). None of the “information ethics” items were identified via “rights” search terms, whereas an unusually high proportion of the “legal precedent” items were identified this way (94%).

The primary questions that are addressed in the identified literature are (1) *Can or could* artificial entities ever be granted moral consideration? (2) *Should* artificial entities be granted moral consideration?[11] The authors use philosophical arguments, ethical arguments, and arguments from legal precedent. They sometimes motivate their arguments with concern for the artificial entities themselves, but others argue in favor of the moral consideration of artificial entities because of positive indirect effects on human society, particularly on moral character (Levy 2009; Davies 2011; Darling 2016, p. 215). Others argue against the moral consideration of artificial entities because of potentially damaging effects on human society (Gerdes 2016; Bryson 2018). Some items, especially those identified via the “moral” search terms, focus on a third question, (3) *What* attitudes do humans currently have vis-a-vis artificial entities, and what predicts these attitudes?[12] A small number of contributions, especially those identified via the “suffering” search terms, also explicitly discuss (4) *What* are the best approaches to ensuring that the suffering of artificial sentience is minimized or that other interests of artificial entities are protected (e.g. Gloor 2016b; Ashrafian 2015a)? Others ask (5) *Should* humanity avoid creating machines that are complex or intelligent enough that they warrant moral consideration (e.g. Basl 2013a; Beckers 2018; Bryson 2018; Hanák 2019; McLaughlin and Rose 2018; Tomasik 2013; Johnson and Verdicchio 2018)?

[9] Bostrom (2014b) has 2,513 citations and was identified via the “suffering” search terms.
[10] Effective altruism is the approach of using our time and resources to help others the most (Sentience Institute 2020). On CLR’s affiliation with this community, see Center on Long-Term Risk (2020).
[11] For summaries, see Tavani (2018), Gunkel, (2018a), and Bennett and Daly (2020).
[12] See the section on “Empirical research on attitudes towards the moral consideration of artificial entities.”

# Dismissal of the importance of moral consideration of artificial entities

Calverley's (2011) chapter in a book on *Machine Ethics* opens with the statement that, "[t]o some, the question of whether legal rights should, or even can, be given to machines is absurd on its face. How, they ask, can pieces of metal, silicon, and plastic, have any attributes that would allow society to assign it any rights at all." Referring to his 1988 essay with Phil McNally, Sohail Inayatullah (2001) notes that he received substantial criticism from colleagues for writing about the topic of robot rights:

> Many years ago in the folly of youth, I wrote an article with a colleague titled, 'The Rights of Robots.' It has been the piece of mine most ridiculed. Pakistani colleagues have mocked me saying that Inayatullah is worried about robot rights while we have neither human rights, economic rights or rights to our own language and local culture – we have only the 'right to be brutalized by our leaders' and Western powers. Others have refused to enter in collegial discussions on the future with me as they have been concerned that I will once again bring up the trivial. As noted thinker Hazel Henderson said, I am happy to join this group – an internet listserv – as long as the rights of robots are not discussed.

Some scholars dismiss discussion of the moral consideration of artificial entities as premature or frivolous, a distraction from concerns that they view as more pressing, usually concerns about the near-term consequences of developments in narrow artificial intelligence and social robots. For example, Chopra (2010, pp. 38-40) dismisses "civil rights for robots" as "fanciful" and "the stuff of good, bad, and simplistic science fiction" but nevertheless argues for the granting of "the status of a legal agent" to computer programs to protect "those that employ and interact with them." Birhane and van Dijk (2020) argue that, "the 'robot rights' debate is focused on first world problems, at the expense of urgent ethical concerns, such as machine bias, machine elicited human labour exploitation, and erosion of privacy all impacting society's least privileged individuals." Cappuccio et al. (2020, p. 3) suggest that arguments in favor of moral consideration for artificial entities that refer to "objective qualities or features, such as freedom of will or sentience" are "problematic because existing social robots are too unsophisticated to be considered sentient"; robots "do not display—and will hardly acquire any time soon—any of the objective cognitive prerequisites that could possibly identify them as persons or moral patients (e.g., self-awareness, autonomous decision, motivations, preferences)." This resembles critiques offered by Coeckelbergh (2010b) and Gunkel (2018c). McLaughlin and Rose (2018) refer to such "objective qualities" but note that, "[r]obot-rights seem not to be much of an issue" in roboethics "because it seems to be fairly widely assumed in the field that the robots in question will be neither sentient nor genuinely intelligent. We think that's a very sensible assumption, at least for the foreseeable future."

Gunkel (2018a, pp. 33-44) provides a number of other examples of critics arguing that discussion of the moral consideration of artificial entities is "ridiculous," as well as cases where it is "given some brief attention only to be bracketed or carefully excluded as an area that shall not be given further thought" or "included by being pushed to the margins of proper consideration… given some attention, but… deliberately marked as something located on the periphery of what is determined to be really important and worthy of investigation." Even Turner (2019), whose discussion of "Rights for AI" is balanced and nuanced, begins their chapter with the admission that, "[g]ranting rights to robots might sound ridiculous."

Despite these attitudes, our analysis shows that academic discussion of the moral consideration of artificial entities is increasing (see Figure 1). This provides evidence that many scholars believe this topic is worth addressing. Indeed, Ziesche and Yampolskiy (2018, p. 2) have proposed the development and formalization of a field of "AI welfare science." They suggest that, "[t]he research should target both aspects for sentient digital minds not to suffer anymore, but also for sentient and non-sentient digital minds not to cause suffering of other sentient digital minds anymore."

Moreover, these dismissals focus on the short-term moral risks regarding the interests of artificial entities. They do not engage with the long-term moral risks discussed in items identified via the "suffering" search terms. Wright (2019) briefly considers the "longer-term" consequences of granting "constitutional rights" to "advanced robots," noting that doing so might spread resources thinly, but this is one of the only items not identified by the "suffering" search terms that explicitly considers the long-term future.[13]

# Attitudes towards the moral consideration of artificial entities among contributing scholars

We might expect different moral frameworks to have radically different implications for attitudes towards the appropriate treatment of artificial entities. Even where scholars share similar moral frameworks, their overall attitudes sometimes differ due to varying timeframes of evaluation or estimations of the likelihood that artificial entities will develop relevant capacities, among other reasons. For example, many scholars use sentience or consciousness as the key criterion determining whether an artificial entity is worthy of moral consideration, and most of these scholars remain open to the possibility that these entities will indeed become sentient in the future. Bryson et al. (2017) view consciousness as an important criterion but note that, "there is no guarantee or necessity that AI [consciousness] will be developed."

The average consideration score (on a scale of 1 to 5) was 3.8 (standard deviation of 0.86) across the 192 items for which a score was assigned, indicating widespread, albeit not universal, agreement among scholars that at least some artificial entities could warrant moral consideration in the future, if not also the present. Where there is enough data to make meaningful comparisons, there is not much difference in average consideration score by country, academic discipline, or the primary framework or moral schema used (see Appendix A).

However, our search terms will have captured only those scholars who deem the subject worthy of at least a passing mention. Other scholars interested in roboethics who consider the subject so "absurd," "ridiculous," "fanciful," or simply irrelevant to their own work that they do not refer to the relevant literature will not have been identified. One of the most forthright arguments against the moral consideration of artificial entities, Bryson's (2010) article "Robots Should be Slaves," though cited 183 times, was not identified by the searches conducted here because of the terminology used in the article.[14] Similarly, scholars who focus on the moral agency of artificial entities, and are therefore unlikely to have been included in this literature review, may tend to see artificial entities as less deserving of moral consideration. For example, Johnson and Miller (2008) focus their paper primarily on the moral agency of artificial entities; the paper was identified by our searches because they cite Sullins' (2005) brief comment encouraging the "moral consideration" of future artificial entities in order to critique Sullins' reasoning. On the other hand, scholars who consider the subject absurd may be more motivated to write on the subject, to critique it, than scholars who do not.

Individuals in disciplines associated with technical research on AI and robotics may be, on average, more hostile to granting moral consideration to artificial entities than researchers from other disciplines. We found that computer engineers and computer scientists had a lower average consideration score than other disciplines (2.6). Additionally, there are many roboticist and AI researcher signatories of the "Open Letter to the European Commission Artificial Intelligence and Robotics" (2018), which objects to a proposal of legal personhood for artificial entities, and when discussion of robot rights has gained media attention, many of the vocal critics appear to have been associated with computer engineering or robotics (Randerson 2007; Kim 2010; Gunkel 2018a, pp. 35-6). Relatedly, Zhang and Dafoe (2019) found in their US survey that respondents with computer science or engineering degrees "rate all AI governance challenges as less important" than other respondents. In this sense, resistance to the moral consideration of artificial entities may fall under a general category of "AI governance" or "AI ethics," which technical researchers may see as less important than other stakeholders. These technical researchers may not disagree with the proponents of moral consideration of artificial entities; they may simply have a different focus, such as incremental technological progress rather than theorizing about societal trajectories.

[13] Torrance (2008), Rademeyer (2017), and Laukyte (2017) also look "a bit further into the future" than other scholars to explore possible causes and effects of granting artificial entities legal rights.

[14] Although "moral agency" is discussed, "moral patiency" is not. The term "rights" is only used in the context of a sentence about "the rights of agency" and when criticizing "animal 'rights' debates."

## Empirical research on attitudes towards the moral consideration of artificial entities

Five papers (Hughes 2005, Nakada 2011, Nakada 2012, Spence et al. 2018, and Lima et al. 2020) included surveys testing whether individuals believe that artificial entities might plausibly warrant moral consideration in the future. Agreement with statements favorable to future moral consideration varied from 9.4% to 70%; given the variety of question wordings, participant nationalities, and sampling methods (students, online participants, or members of the World Transhumanist Association), general trends are difficult to extract.

There are a number of surveys and experiments on attitudes towards current artificial entities. Some of this research provides evidence that people empathize with artificial entities and respond negatively to actions that appear to harm or insult them (Freier 2008; Rosenthal-von der Pütten et al. 2013; Suzuki et al. 2015; Darling 2016). Bartneck and Keijsers (2020) found no significant difference between participants' ratings of the moral acceptability of abuse towards a human or a robot, but other researchers have found evidence that current artificial entities are granted less moral consideration than humans (Slater et al. 2006; Gray et al. 2007; Bartneck and Hu 2008; Küster and Świderska 2016; Akechi et al. 2018; Sommer et al. 2019; Nijssen et al. 2019; Küster and Świderska 2020).

Studies have found that people are more willing to grant artificial entities moral consideration when they have humanlike appearance (Nijssen et al. 2019; Küster et al. 2020), have high emotional (Nijssen et al. 2019 and Lee et al. 2019) or mental capacities (Gray and Wegner 2012; Piazza et al. 2014; Sommer et al. 2019; Nijssen et al. 2019), verbally respond to harm inflicted on them (Freier 2008), or seem to act autonomously (Chernyak and Gary 2016). There is also evidence that people in individual rather than group settings (Hall 2005), with prior experience interacting with robots (Spence et al. 2018), or presented with information promoting support for robot rights, such as "examples of non-human entities that are currently granted legal personhood" (Lima et al. 2020) are more willing to grant artificial entities moral consideration. Other studies have examined the conditions under which people are most willing to attribute high mental capacities to artificial entities (Gray and Wegner 2012; Ward et al. 2013; Briggs et al. 2014; Fraune et al. 2017; McLaughlin and Rose 2018; Wang and Krumhuber 2018; Wortham 2018; Swiderska and Küster 2018, 2020; Küster and Swiderska 2020; Wallkötter et al. 2020; Küster et al. 2020).

# Concluding remarks

Many scholars lament that the moral consideration of artificial entities is discussed infrequently and not viewed as a proper object of academic inquiry. This literature review suggests that these perceptions are no longer entirely accurate. The number of publications is growing exponentially, and most scholars view artificial entities as potentially warranting moral consideration. Still, there are important gaps remaining, suggesting promising opportunities for further research, and the field remains small overall with only 294 items identified in this review.

These discussions have taken place largely separately from each other: legal rights, moral consideration, empirical research on human attitudes, and theoretical exploration of the risks of astronomical suffering among future artificial entities. Further contributions should seek to better integrate these discussions. The analytical frameworks used in one topic may offer valuable contributions to another. For example, what do legal precedent and empirical psychological research suggest are the most likely outcomes for future artificial sentience? What do virtue ethics and rights theories suggest is desirable in these plausible future scenarios?

Despite interest in the topic from policy-makers and the public, there is a notable lack of empirical data about attitudes towards the moral consideration of artificial entities. This leaves scope for surveys and focus groups on a far wider range of predictors of attitudes, experiments that test the effect of various messages and content on these attitudes, and qualitative and computational text analysis of news articles, opinion pieces, and science fiction books and films that touch on these topics. There are also many theoretically interesting questions to be asked about how these attitudes relate to other facets of human society, such as human in-group-out-group and human-animal interactions.

# Acknowledgements

Many thanks to Tobias Baumann, Brian Tomasik, Roman Yampolskiy, Nick Bostrom, Sean Richardson, and Kaj Sotala for providing feedback on earlier drafts of this article.

## Funding

No specific financial support was received for this article.

# Data availability

The datasets generated during and/or analyzed during the current study are available in Appendix A.